\documentclass[10pt,twocolumn,aps,prl,groupedaddress,superscriptaddress,showpacs,noeprint,english,floatfix]{revtex4-2}

\usepackage[T1]{fontenc}
\usepackage[english]{babel}
\usepackage{times}
\usepackage{graphicx}
\usepackage{dcolumn}
\usepackage{bm}
\usepackage{amsmath}
\usepackage{mathrsfs}
\usepackage{amsbsy}
\usepackage{amstext}
\usepackage{amssymb}
\usepackage{setspace}
\usepackage{esint}
\usepackage{xcolor,colortbl}
\usepackage{algpseudocode}
\usepackage[ruled,vlined]{algorithm2e}
\usepackage{float}

\usepackage{relsize}
\usepackage{tikz}

\usepackage[colorlinks,citecolor=blue,urlcolor=blue,linkcolor=blue,hypertexnames=true]{hyperref}
\usepackage{nameref}
\usepackage{orcidlink}
\begin{document}
\title{Scalable Quantum Walk--Based Heuristics for the Minimum Vertex Cover Problem}
\author{Fabr\'icio Souza Luiz \orcidlink{0000-0002-6375-0939}}
\thanks{Corresponding author: fsluiz@unicamp.br}
\affiliation{Instituto de F\'\i sica Gleb Wataghin, Universidade Estadual de Campinas, 13083-859, Campinas, SP, Brazil
}
\author{Arthur Kenzo Feltrin Iwakami
\orcidlink{0000-0001-8888-2367}}\affiliation{Instituto de F\'\i sica Gleb Wataghin, Universidade Estadual de Campinas, 13083-859, Campinas, SP, Brazil
}
\author{Daniel Haro Moraes\orcidlink{0000-0001-8081-8838}}
\affiliation{Venturus Innovation Center, Campinas, SP, Brazil
}
\author{Marcos C\'esar de Oliveira
\orcidlink{0000-0003-2251-2632}}
\affiliation{Instituto de F\'\i sica Gleb Wataghin, Universidade Estadual de Campinas, 13083-859, Campinas, SP, Brazil
}
\affiliation{QuaTI - Quantum Technology \& Information, 13560-161, S\~ao Carlos, SP, Brazil}
\thanks{Corresponding author: marcos@ifi.unicamp.br}
\begin{abstract}
We show that continuous-time quantum walk (CTQW) dynamics on graphs naturally yield a high-quality vertex-selection criterion for the Minimum Vertex Cover (MVC) problem. A greedy algorithm guided by the transition probabilities $P(m)=1-|[e^{i\Gamma t}]_{mm}|^{2}$ of the normalised adjacency Hamiltonian, combined with an energy-penalty freezing mechanism that isolates already-selected vertices, produces near-optimal covers across diverse graph topologies. We implement the algorithm on IBM quantum hardware (\texttt{ibm\_marrakesh}) using only $\lceil\log_{2} V\rceil$ qubits via binary encoding, and on a neutral-atom platform (Bloqade) with one atom per vertex. A short-time expansion reveals that the quantum criterion reduces to the spectral quantity $[\Gamma^{2}]_{mm}=(1/d_{m})\sum_{j\in\mathcal{N}(m)}1/d_{j}$, computable classically in $\mathcal{O}(|E|)$ time per iteration. The resulting quantum-inspired spectral greedy heuristic reproduces the CTQW covers on $98.3\,\%$ of instances across $N \in [4,199]$ (Wilcoxon $p=0.60$; exact MILP reference). Over $14{,}680$ benchmark instances spanning Erd\H{o}s--R\'enyi, Barab\'asi--Albert, and regular graph ensembles ($N \in [4,150]$, exact MILP reference throughout), both algorithms achieve a mean approximation ratio of $1.015$, compared with $1.023$ for degree-greedy, with a worst-case ratio of $1.188$ on random graphs, well below the inapproximability threshold of $\approx 1.3606$~\cite{dinur2005}. At large scale ($V \in [10^3, 10^5]$, $210$ instances), the spectral greedy achieves the best cover in $100\,\%$ of instances, with degree-greedy and Simulated Annealing averaging $1.047$ and $1.072$ times the spectral cover size, respectively.
\end{abstract}
\maketitle
\section*{Introduction}

\begin{figure*}[!ht]
    \centering
    \includegraphics[width=1\linewidth]{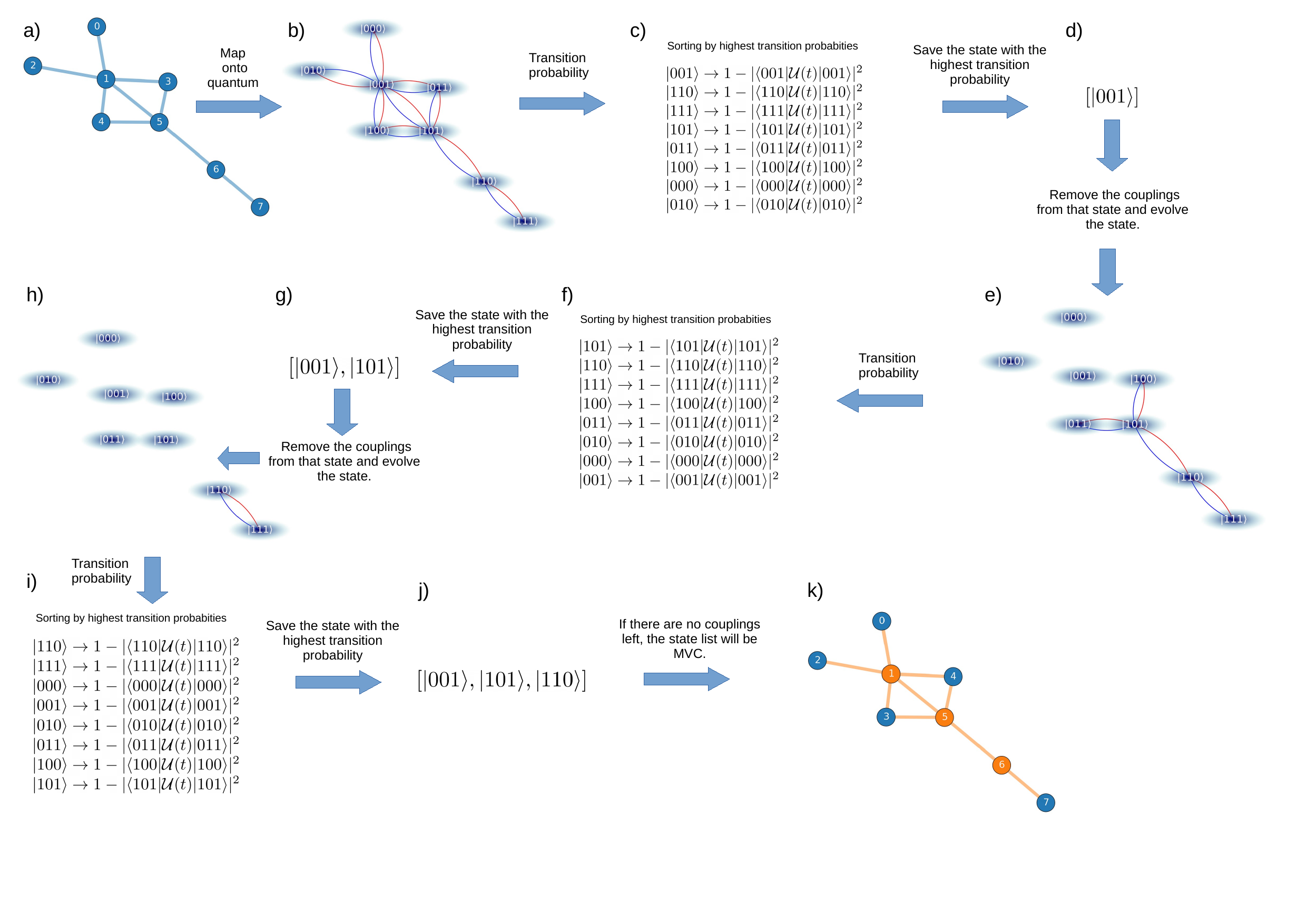}
    \caption{ Iterative Quantum Walk Protocol for MVC. The figure illustrates the sequential algorithm for determining the MVC. (a) An initial graph is defined. (b) The graph is mapped onto quantum states using binary encoding, and the system is evolved via a Quantum Walk. (c-d) Transition probabilities between states are calculated, and the state with the highest probability is identified and saved. (e-g) Iterative Decoupling: The saved state is decoupled from the remaining system, which is then evolved again. The next state with the highest transition probability is selected and saved. (h-j) This decoupling and selection process continues iteratively, identifying the next most probable state. (k) Final Solution: The iteration stops when all remaining states are decoupled (i.e., no more couplings exist between them). The collection of all saved states represents the MVC for the initial graph.}
    \label{fig_expl_pro}
\end{figure*}

The Minimum Vertex Cover (MVC) problem is a central challenge in graph theory and combinatorial optimization, with broad relevance in disciplines such as communication networks, bioinformatics, infrastructure planning, epidemiological modeling, and circuit design \cite{Karp1972,flum2006}. Given a graph $G=(V,E)$, the task is to determine a minimum-size subset of vertices $C \subseteq V$ such that every edge $e \in E$ is incident to at least one vertex in $C$. Despite its simple formulation, the MVC problem is NP-complete~\cite{Karp1972}.

Exact algorithms achieve runtimes $2^{O(k)} \cdot n^{O(1)}$ when cover size $k$ is small \cite{cygan2015parameterized,flum2006}. Classical approximation strategies include the maximal-matching construction (ratio $\le 2$) \cite{papadimitriou1998,hochbaum1997approximation} and strong inapproximability: Dinur and Safra proved no polynomial-time algorithm achieves ratio below $\approx 1.3606$ for large-degree graphs \cite{dinur2005}. These limitations motivate new algorithmic paradigms.

In recent years, quantum algorithms for combinatorial optimization (QAOA \cite{Farhi2022QuantumApproximate,BLEKOS20241,Hadfield2019}, Quantum Alternating Operator Ansatz, and Transverse Vertex Cover formulations \cite{cook2019,bravyi2017}) have been explored, yet still face barriers from hardware noise, circuit depth, and qubit requirements in the Noisy Intermediate-Scale Quantum (NISQ) regime \cite{Preskill2018,Chen2023,Lau2022}. Within this context, Continuous-Time Quantum Walks (CTQWs) have emerged as a powerful framework governed by a Hamiltonian related to the graph Laplacian, exploiting interference, superposition, and coherence \cite{Farhi1998,Childs2009,Venegas-Andraca2012,Kempe2003}. In prior work we demonstrated CTQWs for the Minimum Spanning Tree problem, where vertices with large transition probabilities naturally emerged as structurally relevant \cite{Luiz2025}.

In the context of MVC, a vertex that covers many edges is one from which quantum amplitude spreads most efficiently through the network. The transition probability $P(m\!\to\!\text{out}) = 1 - |[e^{i\Gamma t}]_{mm}|^2$ measures how much amplitude has left site $m$ after time $t$ under the graph Hamiltonian: a large value signals strong coupling to many neighbours, exactly the property that makes $m$ a high-value cover element. This yields a principled physical connection between quantum transport dynamics and vertex cover membership.

In this work we introduce a CTQW-based heuristic for MVC. The algorithm operates iteratively: the graph is encoded into a Hamiltonian via its normalized Laplacian, the quantum system evolves over a short optimal time, and the vertex with the highest transition probability is selected. A spectral isolation (``freezing'') mechanism removes selected vertices from subsequent evolution, systematically constructing a cover while preserving the remaining graph structure.

A key feature is resource efficiency: binary encoding of vertex indices requires only $\lceil \log_2(V) \rceil$ qubits for $V$ vertices. The practical limitation is that the graph Hamiltonian is a dense $V\!\times\!V$ matrix, requiring $\mathcal{O}(V^2)$ two-qubit gates after transpilation---circuit depths that overwhelm current NISQ devices beyond $V \sim 16$ (Table~\ref{tab:qiskit_circuits}). Despite this, binary encodings have received comparatively little attention in the quantum optimization literature~\cite{Farhi2022QuantumApproximate,Hadfield2019}. The central insight of this work is that large-scale quantum hardware is not required: the quantum walk analysis serves as a \emph{discovery mechanism}, revealing a classical selection criterion computable in $\mathcal{O}(|E|)$ time that inherits the quality of the quantum computation.

We benchmark the heuristic against exact MILP solutions and established classical heuristics (Simulated Annealing, FastVC, and the 2-Approximation algorithm) on Erd\H{o}s--R\'enyi, Barab\'asi--Albert, and Regular graph ensembles. Our analysis reveals that the CTQW-based heuristic consistently achieves superior approximation ratios and maintains remarkable robustness with respect to network topology.
We first construct the quantum algorithm and implement it on real hardware; we then show analytically that in the short-time regime the quantum criterion reduces to an efficient classical formula, leading to a quantum-inspired heuristic scalable to graphs with $V \sim 10^6$ vertices. While the gate-based implementation is currently limited to $V \lesssim 16$ due to $\mathcal{O}(V^2)$ circuit depth, the quantum walk analysis reveals a classical criterion whose near-optimal performance across graph topologies would not have been anticipated from classical graph theory alone.

\section*{The CTQW-MVC Quantum Algorithm}

To minimize quantum resources, we encode graph vertices using a binary representation, requiring only $\lceil \log_{2}(V)\rceil$ qubits for $V$ vertices. The edges are mapped to transitions between encoded quantum states (Fig.~\ref{fig_expl_pro}a,b).

We map the system Hamiltonian directly to the normalized symmetric Laplacian $\mathcal{H} = \mathbf{I} - D^{-1/2}AD^{-1/2}$, which provides uniform propagation dynamics across the graph \cite{brouwer2012spectra,chungspectral}, reducing bias toward high-degree vertices. This can be written as
\begin{equation}
    \mathbf{\mathcal{H}} = \sum_{i}\vert i\rangle\langle i\vert - \sum_{i,j}\frac{A_{ij}}{\sqrt{D_{ii}D_{jj}}}\vert i\rangle\langle j\vert,
\end{equation}
or equivalently as $\mathcal{H}\equiv \gamma I-\epsilon \Gamma$ with $\Gamma_{ij}=A_{ij}/\sqrt{D_{ii}D_{jj}}$ and $\epsilon=\gamma=1$ (natural units).

\begin{figure*}[htbp]
    \centering
    \includegraphics[width=\textwidth]{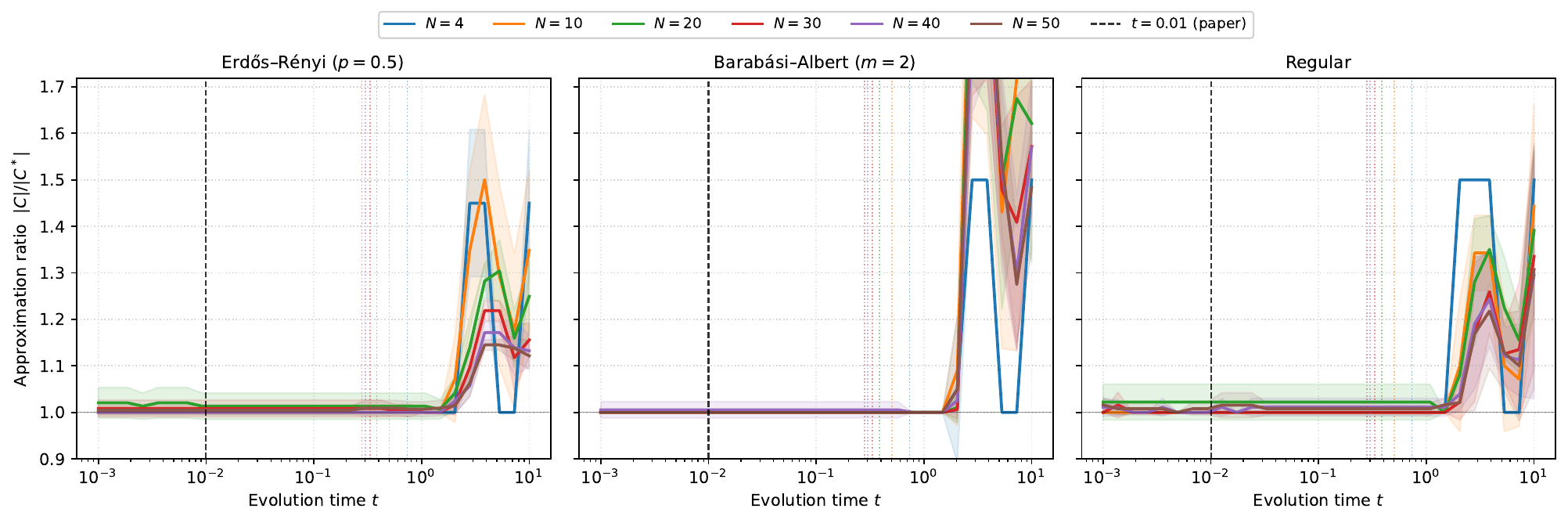}
    \caption{Approximation ratio $|C|/|C^*|$ as a function of evolution time $t$ (log scale) for CTQW on ER ($p=0.5$), BA ($m=2$), and Regular graphs. Each curve corresponds to a fixed graph size $N$ (color-coded); shaded bands show $\pm 1\sigma$ over $10$ graph instances. Vertical dashed line: $t=0.01$; dotted vertical lines: theoretical $t_\mathrm{opt}(N)$ for each $N$.}
    \label{fig:t_sensitivity}
\end{figure*}

The evolution operator is $U(t)=e^{-\imath\mathcal{H}t}$ ($\hbar=1$). We employ an optimal evolution time $t_{\rm opt}$ in the coherent transport window \cite{Muelken2011}: long enough to probe local connectivity, yet short enough to avoid destructive self-interference. Based on prior work \cite{Luiz2025,ChildsGoldstone2004}:
\begin{equation}
    t_{\rm opt} = \frac{4}{\pi \sqrt{V}} + 0.1.
\end{equation}
The sensitivity of algorithm quality to the precise choice of $t$ is assessed in Fig.~\ref{fig:t_sensitivity}; $t = 0.01$ lies in a broad plateau of near-optimal performance for all graph sizes tested.

Since $\mathcal{H} = \mathbf{I} - \Gamma$ and the identity commutes with $\Gamma$, the evolution factors exactly as $\mathcal{U}(t) = e^{-it}\,e^{i\Gamma t}$. The global phase cancels in the probability, yielding the transition probability:
\begin{eqnarray}
    P(m\rightarrow \text{out}) &=& 1-\left\vert \left(e^{i t \Gamma}\right)_{mm}\right\vert^{2},\label{prob}
\end{eqnarray}
where $\Gamma_{ij} = A_{ij}/\sqrt{D_{ii}D_{jj}}$ and the subscript $mm$ denotes the $m$-th diagonal element.

Our heuristic is an iterative process that requires removing a selected vertex and its incident edges from the graph after each step. Classically, this is achieved by zeroing the corresponding entries of the adjacency matrix ($A_{j,m}=A_{m,j}=0$ for all neighbours $j$), producing a modified Hamiltonian for the next iteration. For implementation on a quantum processor, we employ a spectral isolation (``freezing'') mechanism~\cite{Castelano2024}: the total Hamiltonian is decomposed as
\begin{equation}
    \mathcal{H}_{\text{total}} = \mathcal{H}_{j} + \mathcal{H}_{m} + \mathcal{H}_{F},
\end{equation}
where $\mathcal{H}_{j} = P_j \mathcal{H} P_j$ acts on the uncovered subspace, $\mathcal{H}_{m} = P_m \mathcal{H} P_m$ acts on already-selected vertices, and $\mathcal{H}_{F} = (\Omega-1) \cdot P_m$ applies a large energy penalty that shifts the selected-vertex subspace far off-resonance from the active dynamics. In the interaction picture, for $\Omega \gg \|\mathcal{H}\|$, the rapid phase oscillation of the frozen subspace averages out its coupling to the active vertices, and the effective evolution reduces to:
\[ i \frac{\partial\vert\psi_{I,j}(t)\rangle}{\partial t} \approx  \mathcal{H}_{j}  \vert\psi_{I,j}(t)\rangle.\]

The leakage probability into the frozen subspace satisfies $P_{\rm leak}(t) \le \min(t^{2},\,4/(\Omega-3)^2)$ (derived in Appendix~\ref{app:derivation}); for $t = 0.01$ and $\Omega \ge 10$, $P_{\rm leak} \le 10^{-4}$, so the active-subspace approximation holds to better than $0.01\%$ per step.

The CTQW-MVC heuristic is summarised in Algorithm~\ref{alg:ctqw}. On quantum hardware, Step~3 is executed by state preparation and measurement; on classical hardware, by direct computation of $U_{mm}(t)$ via matrix exponentiation.

\begin{algorithm}[H]
\caption{CTQW-MVC heuristic}
\label{alg:ctqw}
\KwIn{Graph $G=(V,E)$, evolution time $t$ (default: $t=0.01$)}
\KwOut{Vertex cover $C$}
Construct $\mathcal{H} = \mathbf{I} - D^{-1/2}AD^{-1/2}$\;
Initialize $C \leftarrow \emptyset$\;
\While{$E \neq \emptyset$}{
    Compute $U(t) = e^{i\Gamma t}$ \quad [quantum: prepare $|m\rangle$, evolve, measure; classical: matrix exponentiation]\;
    \ForEach{active vertex $m$ ($\deg(m)>0$)}{
        $P(m\!\rightarrow\!\text{out}) \leftarrow 1 - |U_{mm}|^2$\;
    }
    $m^* \leftarrow \arg\max_{m} P(m\!\rightarrow\!\text{out})$\;
    Add $m^*$ to $C$; remove all edges incident to $m^*$ from $E$\;
}
\Return $C$\;
\end{algorithm}

As a greedy procedure, the algorithm does not perform backtracking. The CTQW dynamics mitigate suboptimality: transition probabilities $P(m\!\rightarrow\!\text{out})$ reflect the full spectral structure of the normalised Laplacian, distributing selection weight more uniformly than a simple maximum-degree greedy rule. Ties in $\arg\max$ are broken by \texttt{numpy.argmax} (lowest-index vertex); this choice is inconsequential: over 26 regular-graph instances (the topology most susceptible to ties), 50 random relabellings per graph produced identical cover sizes in all cases (standard deviation $=0$ across all 1\,300 runs). Extensions (randomised restarts, post-processing pruning, hybrid local search) are discussed in the Limitations section.

The per-iteration cost is dominated by computing $e^{-i\mathcal{H}t}$ of the $V\!\times\!V$ normalised Laplacian: $\mathcal{O}(V^3)$ using Pad\'e-approximant methods \cite{Hatano2005}. The algorithm terminates in at most $|C|$ steps, giving total complexity $\mathcal{O}(|C|\cdot V^3)$ and space $\mathcal{O}(V^2)$. For sparse graphs with $|E| = \mathcal{O}(V)$, Krylov-subspace methods (\texttt{expm\_multiply}) compute $e^{i\Gamma t}|m\rangle$ in $\mathcal{O}(k\,|E|)$ per Krylov dimension $k$; since the diagonal element $[e^{i\Gamma t}]_{mm}$ requires one such call per vertex, the total per-iteration cost is $\mathcal{O}(k V |E|) = \mathcal{O}(V^2)$ for $|E|=\mathcal{O}(V)$. In practice, Algorithm~\ref{alg:spectral} supersedes this approach for sparse graphs, evaluating $[\Gamma^2]_{mm}$ in $\mathcal{O}(|E|)$ without any quantum simulation. The binary encoding keeps the simulation space $V$-dimensional, unlike one-hot encodings which operate in $2^V$-dimensional Hilbert space (classically intractable beyond $V\sim 50$). Table~\ref{tab:runtime} reports measured wall-clock times across a representative subset of instances ($V \in \{4,\ldots,54\}$).

\begin{table}[htbp]
\centering
\caption{Mean wall-clock time per full run (ms) for $V$ vertices, averaged
over all graph types and seeds in the benchmark. CTQW times reflect
classical scipy simulation (\texttt{expm\_multiply}); on quantum hardware
the evolution step scales as $\mathcal{O}(1)$ per iteration.}
\label{tab:runtime}
\begin{tabular}{cccccc}
\hline
$V$ & CTQW & Deg.-Greedy & 2-Approx & FastVC & Sim.\ Ann. \\
\hline
 9  &  15.4 & 0.11 & 0.08 & 0.08 &   6.1 \\
19  &  57.5 & 0.27 & 0.16 & 0.21 &  13.5 \\
29  & 123.0 & 0.53 & 0.30 & 0.43 &  23.5 \\
39  & 212.6 & 0.83 & 0.49 & 0.70 &  36.2 \\
54  & 407.3 & 1.43 & 0.83 & 1.35 &  60.9 \\
\hline
\end{tabular}
\end{table}

CTQW-MVC is a genuinely quantum algorithm: the selection criterion $P(m\!\to\!\text{out}) = 1 - |\langle m|U(t)|m\rangle|^2$ results from coherent interference under the graph Hamiltonian. The binary encoding stores $V$ vertex indices in $n = \lceil\log_2 V\rceil$ qubits.

\paragraph{Gate-based implementation (IBM Qiskit).}
We implement CTQW-MVC as a Qiskit circuit: build the $V\!\times\!V$ normalised Laplacian, pad it to $2^n \times 2^n$, compute $U(t)$ via exact matrix exponentiation, and embed as a \texttt{UnitaryGate} on $n$ qubits.

\begin{figure*}[!t]
    \centering
    \includegraphics[width=\textwidth]{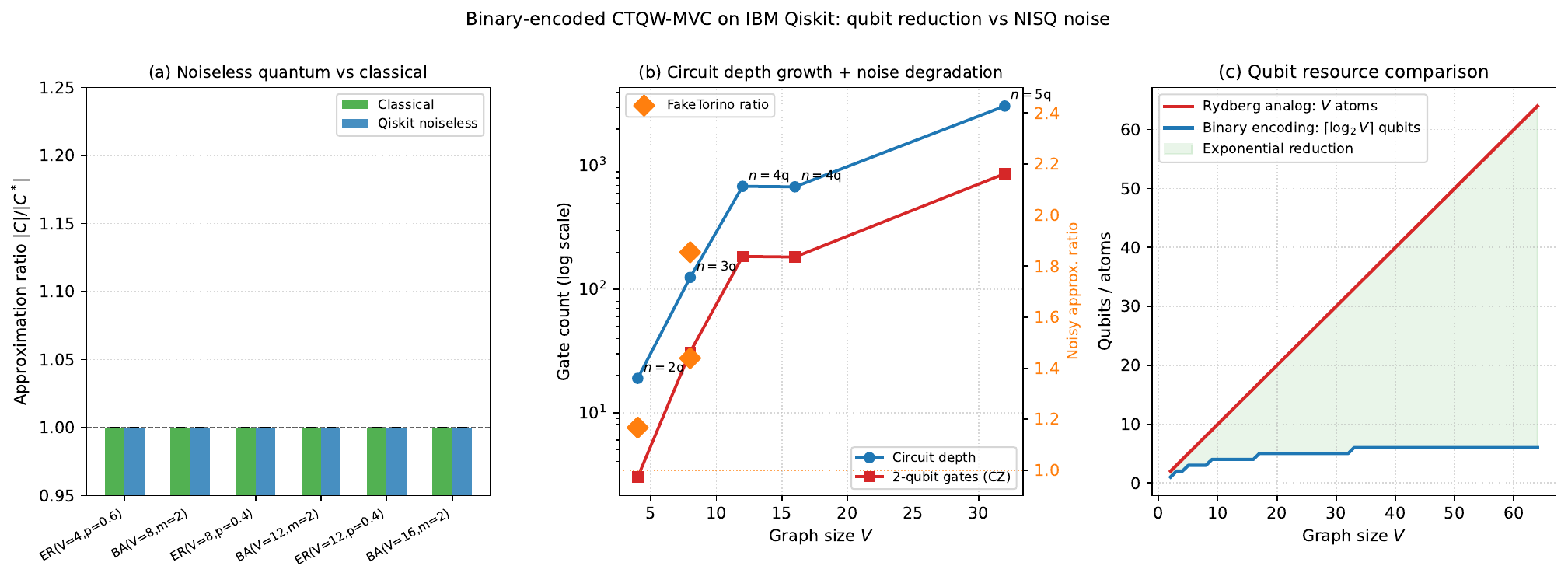}
    \caption{Gate-based CTQW-MVC on IBM Qiskit with binary encoding ($\lceil\log_2 V\rceil$ qubits).
    \textbf{(a)} Approximation ratio for classical scipy (green) and noiseless Qiskit statevector (blue) across ER and BA graphs at $V \in \{4,8,12,16\}$ (38 instances); noiseless quantum matches classical, confirming the algorithm computes genuine quantum transition probabilities.
    \textbf{(b)} Circuit depth and CZ gate count after transpilation to FakeTorino native gates (log scale); FakeTorino noise-model ratios shown for $V\in\{4,8\}$.
    \textbf{(c)} Qubit resource comparison: binary encoding ($\lceil\log_2 V\rceil$, blue) vs.\ Rydberg analog ($V$ atoms, red), highlighting the exponential advantage for fault-tolerant hardware.}
    \label{fig:qiskit_results}
\end{figure*}

The noiseless Qiskit statevector simulator on $V \in \{4, 8, 12, 16\}$ graphs (38 instances, Fig.~\ref{fig:qiskit_results}a) produces identical approximation ratios to the classical scipy reference, confirming that the algorithm computes genuine quantum transition probabilities---not a classical approximation.

The Hamiltonian $\mathcal{H}$ is a dense non-local matrix requiring $\mathcal{O}(4^n) \sim \mathcal{O}(V^2)$ two-qubit gate terms, so transpiled circuit depth grows rapidly with $V$ (Table~\ref{tab:qiskit_circuits}).

\begin{table}[htbp]
\centering
\caption{Circuit cost for one CTQW step after transpilation to ibm\_marrakesh
native gates (optimisation level 1, heavy-hex topology, basis: $\{R_z, SX, CZ\}$).
The qubit count grows as $\lceil\log_2 V\rceil$; circuit depth and two-qubit gate
count grow as $\mathcal{O}(V^2)$.}
\label{tab:qiskit_circuits}
\begin{tabular}{ccccc}
\hline
$V$ & $\lceil\log_2 V\rceil$ & Depth & 2-qubit gates & HW ratio (mean) \\
\hline
 4 & 2 &    23 &     3 & $1.000$ \\
 8 & 3 &   146 &    37 & $1.750$ \\
16 & 4 &   806 &   245 & $1.191$ \\
32 & 5 &  3445 &  1044 & $1.857$ \\
\hline
\end{tabular}
\end{table}

We executed CTQW-MVC on \texttt{ibm\_marrakesh} (156-qubit heavy-hex processor) for $V \in \{4, 8, 16, 32\}$, using \texttt{SamplerV2} with $256$ shots per circuit. At $V = 4$ ($n = 2$, depth $= 23$) the hardware solution is exact: ratio $= 1.000$ on all instances. At $V = 32$ ($n = 5$, depth $= 3445$, 1044 CZ gates), accumulated gate errors dominate: ratio $1.857$, consistent with fidelity $0.995^{1044} \approx 0.005$. All hardware solutions are valid vertex covers in every trial, confirming the NISQ bottleneck is circuit depth, not qubit count, and the binary encoding is viable for $V \lesssim 16$.
The 256-shot budget deserves consideration: for $n$ qubits there are $2^n$ possible outcomes, yielding on average $256/2^n$ shots per outcome and a shot-noise standard deviation $\sigma \approx \sqrt{p(1-p)/256} \approx 0.03$--$0.11$ per transition probability. For $V=4$ ($\sim\!64$ shots/outcome) and $V=8$ ($\sim\!32$/outcome) the noise is small relative to probability differences. For $V=32$ ($\sim\!8$/outcome, $\sigma \approx 0.11$) shot noise contributes to---but does not solely explain---the ratio degradation; circuit infidelity ($0.995^{1044}\approx 0.005$) is the dominant factor.

\begin{figure*}[!t]
    \centering
    \includegraphics[width=\textwidth]{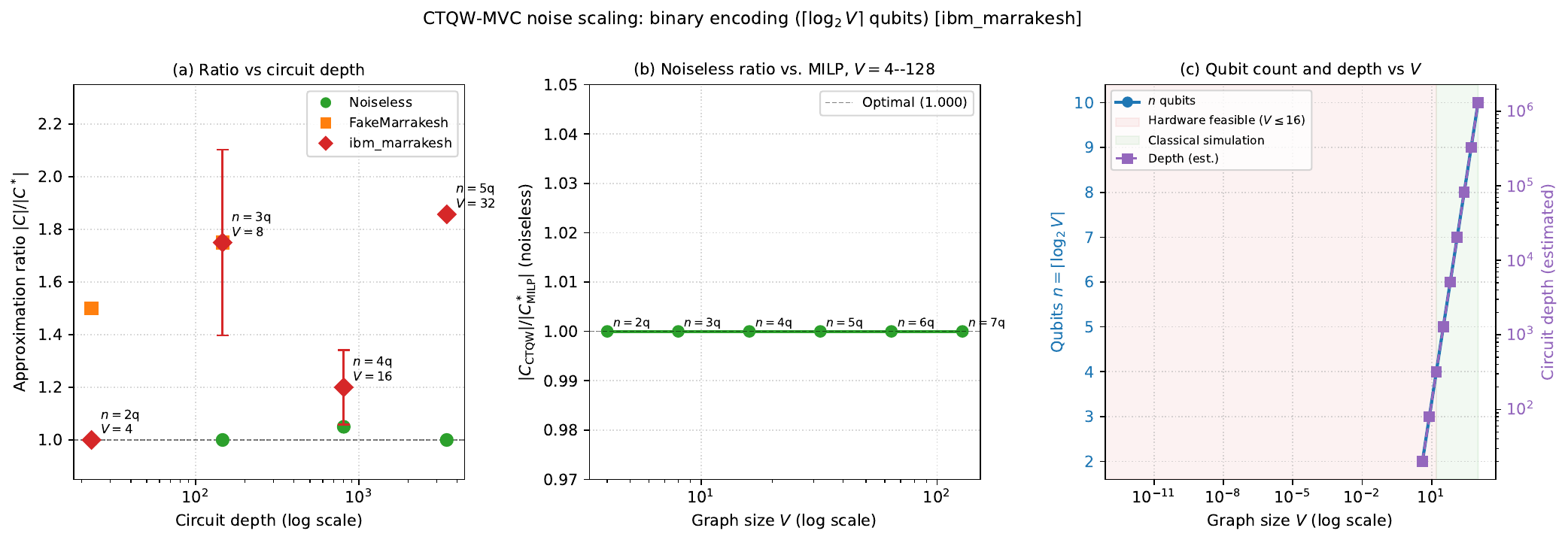}
    \caption{Noise scaling of CTQW-MVC with binary encoding on \texttt{ibm\_marrakesh}.
    \textbf{(a)} Approximation ratio vs.\ transpiled circuit depth for noiseless (green circles), FakeMarrakesh (orange squares), and real hardware (red diamonds) at $V \in \{4,8,16,32\}$.
    \textbf{(b)} Noiseless ratio $|C_{\rm CTQW}|/|C^*_{\rm MILP}|$ vs.\ $V$ for $V \in \{4,8,16,32,64,128\}$ ($6$ instances, one per $V$, exact MILP reference): CTQW finds the MILP-exact optimum (${\rm ratio}=1.000$) in all cases.
    \textbf{(c)} Qubit count $n = \lceil\log_2 V\rceil$ and estimated circuit depth $\propto 4^n$ vs.\ $V$. The shaded region marks the NISQ-viable range ($V \le 16$, depth $\lesssim 10^3$).}
    \label{fig:noise_scaling}
\end{figure*}

\paragraph{Encoding regimes and hardware trade-offs.}
The two hardware paths use qubit resources differently: gate-based binary encoding stores $V$ indices in $\lceil\log_2 V\rceil$ qubits at the cost of $\mathcal{O}(V^2)$ two-qubit gates; neutral-atom analog uses one atom per vertex with no gate decomposition. Gate-based hardware suits the long-term fault-tolerant regime; the analog approach is immediately deployable for unit-disk-embeddable graphs (Appendix~\ref{app:rydberg}).

\section*{From Quantum Dynamics to an Efficient Classical Criterion}
\label{sec:classical_criterion}

The quantum algorithm of the preceding section provides exact transition probabilities through the matrix exponential $e^{i\Gamma t}$. We now show that a systematic short-time expansion reveals the analytical structure behind the algorithm's performance, leading to a classical criterion of equivalent quality but dramatically lower computational cost.

Expanding the matrix exponential $e^{i\Gamma t}$ to second order in $t$ and exploiting $\Gamma_{mm}=0$ (no self-loops), one finds that the
diagonal element of $\Gamma^2$ is (Appendix~\ref{app:derivation}):
\begin{equation}
    [\Gamma^2]_{mm}
      = \frac{1}{d_m}\sum_{j\in\mathcal{N}(m)}\frac{1}{d_j},
    \label{eq:gamma2}
\end{equation}
where $\mathcal{N}(m)$ denotes the neighbourhood of vertex $m$. Since the imaginary part of $[e^{i\Gamma t}]_{mm}$ first appears at
$\mathcal{O}(t^3)$ and contributes only at $\mathcal{O}(t^6)$ to the modulus squared, the transition probability reduces to
\begin{align}
    P(m\!\rightarrow\!\text{out})
    &= t^2[\Gamma^2]_{mm} + \mathcal{O}(t^4)\nonumber\\
    &= \frac{t^2}{d_m}\!\sum_{j\in\mathcal{N}(m)}\!\frac{1}{d_j}
      + \mathcal{O}(t^4).
    \label{eq:short_time}
\end{align}
At short times the criterion favours vertices that connect to many low-degree neighbours---a spectral quantity that captures both local
connectivity and neighbourhood structure, going beyond pure degree counting.  The quantity $[\Gamma^2]_{mm} = (1/d_m)\sum_{j\in\mathcal{N}(m)}1/d_j$ is the average inverse degree of vertex $m$'s neighbours (the mean of $\{1/d_j\}_{j\in\mathcal{N}(m)}$, scaled by $1/d_m$); a two-hop spectral measure that encodes structural information inaccessible to single-step degree-based heuristics.

We use $t = 0.01$ throughout, lying in the regime $t \ll t_{\rm opt}$ for all graph sizes tested.  The complete derivation with all
intermediate steps is given in Appendix~\ref{app:derivation}.

We validated the short-time equivalence across two complementary datasets, both using SageMath's MILP exact solver as reference. In the primary dataset ($N \in [4,\,54]$, $462$ instances spanning Erd\H{o}s--R\'enyi with $p\in\{0.3,0.5,0.7\}$, Barab\'asi--Albert with
$m\in\{2,5\}$, and regular topologies), the two algorithms select identical vertex covers in $457/462$ cases ($98.9\,\%$; $95\,\%$ Wilson CI:
$[97.5\,\%,\,99.5\,\%]$); Erd\H{o}s--R\'enyi agreement is exact ($100\,\%$). In the extended dataset ($N \in [4,\,199]$, $350$ instances, exact MILP reference in $97\,\%$ of cases), equivalence reaches $344/350$ ($98.3\,\%$; $95\,\%$ CI: $[96.3\,\%,\,99.2\,\%]$) with $100\,\%$ agreement on all Erd\H{o}s--R\'enyi instances and, notably, on all $N=199$ instances. (The full benchmark of Sec.~\ref{sec:results} uses a broader parameter sweep totalling $14{,}680$ instances.)

The discordant cases occur exclusively in Barab\'asi--Albert and regular graphs at $N \geq 34$, where vertex scores become degenerate or
quasi-degenerate and selection reduces to numerical tie-breaking (see Appendix~\ref{app:derivation}: spectral scores are identically equal on
regular graphs while CTQW probabilities carry floating-point perturbations of order $\sim10^{-11}$ from evaluation of the matrix exponential in the padded $2^{\lceil\log_2 N\rceil}$ space).  Neither algorithm systematically outperforms the other in these cases; cover-size differences never exceed~$2$.

The pooled approximation ratios are statistically indistinguishable across both datasets: $1.005\pm0.013$ (CTQW) versus $1.005\pm0.013$ (spectral greedy) on the extended set; a two-sided Wilcoxon signed-rank test yields $p=0.60$.  Both substantially outperform degree-greedy selection ($1.025\pm0.037$; $p<10^{-22}$) and Simulated Annealing ($1.026\pm0.032$; $p<10^{-4}$).  The spectral greedy runs in $33\,\mathrm{ms}$ per instance versus $1.52\,\mathrm{s}$ for the full CTQW on the extended set; a $46\times$ speedup that grows with graph size, reaching $\sim2\,\mathrm{s}$ for the CTQW at $N=199$ (exponentiating a $256\times256$ matrix) versus $33\,\mathrm{ms}$ for the spectral greedy, consistent with the $\mathcal{O}(V^3)$ versus $\mathcal{O}(|E|)$ scaling per greedy step.

To assess sensitivity to the choice of $t$, we swept $t$ over a logarithmic grid of $30$ values from $10^{-3}$ to $10^{1}$ on fresh ER ($p=0.5$), BA ($m=2$), and regular ensembles ($N \in \{4,10,20,30,40,50\}$, $10$ instances each). Results are shown in Fig.~\ref{fig:t_sensitivity}.

Three regimes emerge.  For $t \lesssim 1$, near-optimal performance with mean ratios in $[1.000,\,1.022]$---a broad plateau comfortably containing $t = 0.01$. For $t \gtrsim 3$, the ratio degrades noticeably (worst cases: $1.29$ ER, $1.95$ BA, $1.36$ REG at $t = 3.86$).  The intermediate regime $1 \lesssim t \lesssim 3$ is topology-dependent, with BA graphs degrading earlier than ER or regular, consistent with the uneven spectral gap introduced by hub nodes. The agreement between CTQW and spectral greedy reflects this regime structure: it remains at $100\,\%$ for ER and BA graphs throughout $t \in [10^{-3}, 0.1]$, then decreases for $t \gtrsim 1$ (ER drops below $95\,\%$ at $t \approx 1$;
BA at $t \approx 2$), and collapses to near zero at $t = 10$ for all topologies. Regular graphs maintain a constant $\approx80\,\%$ agreement across $t \in [10^{-3}, 1]$, consistent with the tie-breaking origin identified above. These results confirm that equivalence is an intrinsic property of the short-time regime, not a coincidence at a single value.

The theoretical coherent window $t_\mathrm{opt} = 4/(\pi\sqrt{N}) + 0.1$ predicts $t \in [0.28, 0.74]$ for $N \in [4, 50]$, fully contained within the observed plateau, see Fig.(\ref{fig:t_sensitivity}).  The broad plateau $t \in [10^{-3}, 1]$ confirms that the short-time approximation is not a fine-tuned choice but an intrinsic property of the spectral criterion.

The analytical result of Eq.~(\ref{eq:gamma2}) enables a purely classical algorithm that bypasses the matrix exponential entirely. We present this as Algorithm~\ref{alg:spectral}.

\begin{algorithm}[H]
\caption{Quantum-inspired spectral greedy (classical, $O(E)$ per iteration)}
\label{alg:spectral}
\KwIn{Graph $G=(V,E)$}
\KwOut{Vertex cover $C$}
Initialize $C \leftarrow \emptyset$\;
\While{$E \neq \emptyset$}{
    \ForEach{active vertex $m$ ($\deg(m)>0$)}{
        $s(m) \leftarrow \displaystyle\frac{1}{d_m}\sum_{j \in \mathcal{N}(m)} \frac{1}{d_j}$\;
    }
    $m^* \leftarrow \arg\max_{m} s(m)$\;
    Add $m^*$ to $C$; remove all edges incident to $m^*$ from $E$\;
}
\Return $C$\;
\end{algorithm}

Algorithm~\ref{alg:spectral} computes the score $s(m) = [\Gamma^2]_{mm}$ by a single pass over the edge list: for each edge $(m,j)$, add $1/d_j$ to an accumulator for $m$ and $1/d_m$ to an accumulator for $j$; then divide each accumulator by the vertex degree. This requires $\mathcal{O}(|E|)$ operations per iteration, reducing to $\mathcal{O}(V)$ for sparse graphs.

Table~\ref{tab:cost_comparison} summarises the computational cost per iteration for each implementation of the CTQW-MVC criterion.

\begin{table*}[htbp]
\centering
\caption{Computational cost per iteration for each implementation.}
\label{tab:cost_comparison}
\begin{tabular}{lll}
\hline
Implementation & Cost per iteration & Notes \\
\hline
Quantum circuit (gate-based) & $\mathcal{O}(V^2)$ 2-qubit gates & $\lceil\log_2 V\rceil$ qubits \\
Quantum analog (Rydberg) & $\mathcal{O}(1)$ (physical evolution) & $V$ atoms, UDG only \\
Classical CTQW (\texttt{expm}) & $\mathcal{O}(V^3)$ / $\mathcal{O}(V^2)$ & Pad\'e / Krylov \\
\textbf{Spectral greedy} (Alg.~\ref{alg:spectral}) & $\mathbf{\mathcal{O}(|E|)}$ & $\mathcal{O}(V)$ for sparse \\
\hline
\end{tabular}
\end{table*}

Algorithm~\ref{alg:spectral} is the direct practical output of the quantum analysis: a greedy heuristic whose selection criterion was derived from first principles of quantum transport, computable without any quantum hardware at the same cost as degree-greedy.

\section*{Results and Discussion}
\label{sec:results}

The benchmarks presented below validate both Algorithm~\ref{alg:ctqw} (CTQW) and Algorithm~\ref{alg:spectral} (spectral greedy); since the two produce statistically indistinguishable covers on $98.3\,\%$ of instances across $N \in [4,199]$ (Wilcoxon $p=0.60$), the results apply to both implementations.

To assess robustness and generality, we employed three complementary classes of synthetic random graphs: Erd\H{o}s--R\'enyi (ER), Barab\'asi--Albert (BA), and Regular (REG) \cite{Bollobas_Erdös_1976,Karoński1997,Barabási1999,chen1997graph}.

ER graphs: $N \in [4, 154]$, edge probabilities $p \in \{0.2, 0.4, 0.5, 0.6, 0.7, 0.8, 0.9\}$, 10 independent instances per $(N,p)$ configuration.

BA graphs: $N \in [4, 154]$, attachment parameter $m \in \{1, 2, 3, 5, 10, 15\}$, 10 instances per $(N,m)$ configuration.

Regular graphs: $N \in [4, 154]$, degrees $k \in \{2, \lfloor 0.25N\rfloor, \lfloor 0.5N\rfloor\}$, up to 5 non-isomorphic connected instances per $(N,k)$ (verified via Weisfeiler--Lehman hash).

Computational parameters: all edge weights $w_{ij} = 1.0$; fixed evolution time $t = 0.01$; exact reference via MILP (SageMath), available for all instances ($N \le 150$).

Table~\ref{tab:dg_comparison} summarises the mean approximation ratios by topology; a visual heatmap is provided in Appendix~\ref{app:heatmap}.

\paragraph{Comparison with degree-greedy heuristic.}
A natural baseline for topology-aware vertex selection is the degree-greedy heuristic: at each step, the vertex with the highest current degree is added to the cover and removed from the graph. This purely local rule has $O(V \cdot E)$ complexity and is competitive on sparse instances~\cite{Kann1992}.

\begin{figure*}[htbp]
    \centering
    \includegraphics[width=\textwidth]{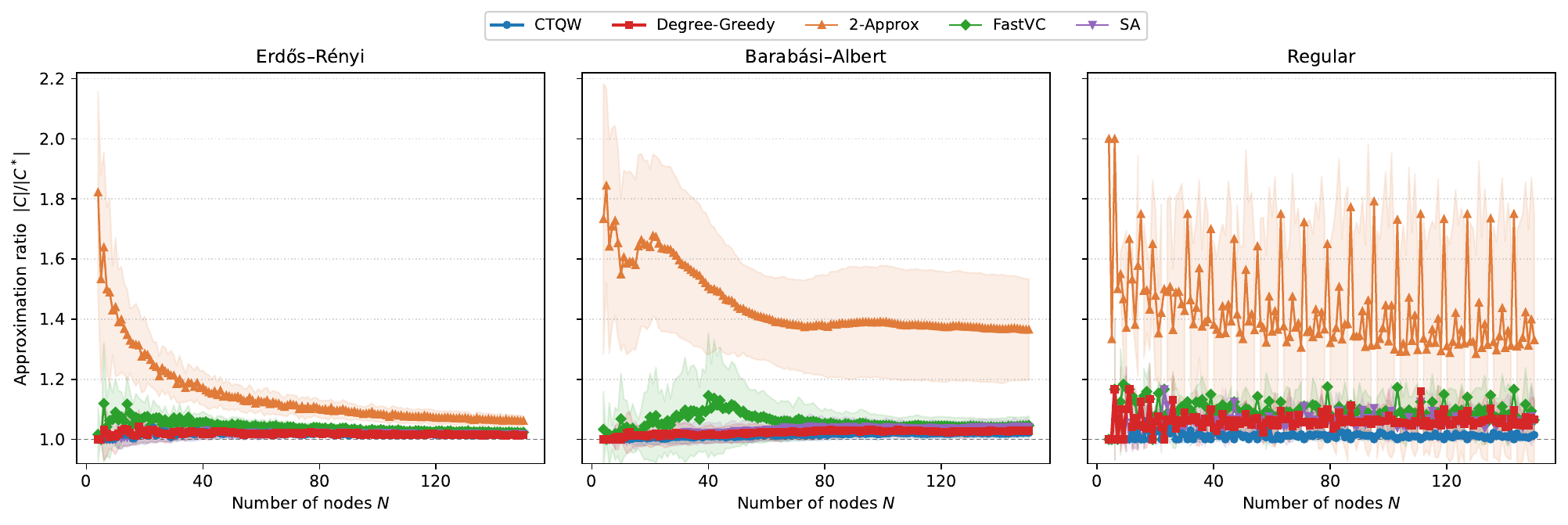}
    \caption{Approximation ratio $|C|/|C^{*}|$ as a function of graph size $N$ for CTQW (blue circles), Degree-Greedy (red squares), 2-Approx (orange triangles), FastVC (green diamonds), and Simulated Annealing (purple inverted triangles). Results averaged over $10$ graph instances per $(N, \text{param})$ combination; shaded bands show $\pm 1\sigma$. Reference sizes use the SageMath MILP exact solver for all $N \le 150$.}
    \label{fig:degree_greedy}
\end{figure*}

Figure~\ref{fig:degree_greedy} compares CTQW against degree-greedy and three classical baselines over ensembles of ER ($p \in \{0.4, 0.5, 0.7\}$), BA ($m \in \{1, 2, 3, 5, 10, 15, 20\}$, valid $m < N$), and Regular graphs spanning $N \in [4, 150]$ (10 instances per $(N,\text{param})$ combination, $14{,}680$ graphs in total). Reference cover sizes were computed using the SageMath MILP exact solver for all sizes.

The results are summarised in Table~\ref{tab:dg_comparison}. Pooled over all topologies, CTQW achieves a mean approximation ratio of $1.015$, compared with $1.023$ for degree-greedy. The advantage is most pronounced on Regular graphs (CTQW $1.013$ vs.\ degree-greedy $1.060$), where spectral degeneracy forces degree-based rules to make essentially arbitrary tie-breaking choices, whereas the CTQW amplitude naturally differentiates structurally equivalent vertices through interference. CTQW produces a strictly smaller cover in $32.6\,\%$ of individual instances, ties with degree-greedy in $51.7\,\%$, and is exceeded in $15.7\,\%$; it is never worse on average for any topology.

These results confirm that the CTQW performance advantage over degree-greedy cannot be attributed solely to greedy degree exploitation: the quantum walk encodes global spectral information of the normalized Laplacian, providing structural correlations beyond single-step degree counts.

To assess statistical significance, we applied a one-sided Wilcoxon signed-rank test~\cite{wilcoxon1945} (null hypothesis: equal cover size; alternative: CTQW produces smaller cover) to all $14{,}680$ paired instances. CTQW is significantly better than every baseline ($p < 10^{-261}$), with effect sizes ranging from $r=0.41$ (vs.\ degree-greedy) to $r=0.87$ (vs.\ 2-Approximation), confirming the advantage is not due to chance (Table~\ref{tab:wilcoxon}).

\begin{table}[htbp]
\centering
\caption{Wilcoxon signed-rank test: CTQW vs.\ each baseline ($n=14{,}680$ paired instances, one-sided alternative CTQW $<$ baseline).  $n_{\neq}$: pairs with unequal cover sizes; $r = |Z|/\sqrt{n_{\neq}}$: effect size.}
\label{tab:wilcoxon}
\begin{tabular}{lrrrr}
\hline
Baseline & $W$ & $p$-value & $n_{\neq}$ & $r$ \\
\hline
Degree-Greedy       & $6.6\times10^{6}$ & $2.3\times10^{-262}$ & 7093  & 0.41 \\
2-Approximation     & $0$               & $<10^{-300}$         & 14647 & 0.87 \\
FastVC              & $2.3\times10^{6}$ & $<10^{-300}$         & 9545  & 0.78 \\
Simulated Annealing & $5.1\times10^{6}$ & $<10^{-300}$         & 7722  & 0.57 \\
\hline
\end{tabular}
\end{table}

\begin{table*}[htbp]
\centering
\caption{Mean approximation ratio $|C|/|C^{*}|$ by graph topology. Lower is better; $1.0$ is optimal. Results averaged over $14{,}680$ graph instances ($N \in [4,150]$, exact MILP reference throughout, $10$ instances per $(N,\text{param})$ combination).}
\label{tab:dg_comparison}
\begin{tabular}{lrrrrr}
\hline
Graph & CTQW & Degree-Greedy & 2-Approx & FastVC & SA \\
\hline
Erd\H{o}s--R\'enyi  & $1.017$ & $1.019$ & $1.149$ & $1.041$ & $1.018$ \\
Barab\'asi--Albert & $1.015$ & $1.023$ & $1.441$ & $1.058$ & $1.031$ \\
Regular      & $1.013$ & $1.060$ & $1.399$ & $1.090$ & $1.059$ \\
\hline
\textbf{Pooled} & $\mathbf{1.015}$ & $\mathbf{1.023}$ & $\mathbf{1.352}$ & $\mathbf{1.054}$ & $\mathbf{1.027}$ \\
\hline
\end{tabular}
\end{table*}

In summary, while classical heuristics exhibit clear dependencies on structural features (randomness in ER, hub dominance in BA, and spectral uniformity in REG) the CTQW-based quantum heuristic remains largely invariant to these effects, arising from the intrinsic ability of quantum evolution to sample and correlate the network structure in a balanced and topology-agnostic manner.

\paragraph{Comparison with variational quantum approaches.}
The Quantum Approximate Optimization Algorithm (QAOA)~\cite{Farhi2014} and its constrained generalizations~\cite{Hadfield2019} represent the leading variational framework for combinatorial problems on near-term hardware. For MVC, standard QAOA formulations encode the problem with one qubit per vertex and enforce the coverage constraint via penalty terms in the cost Hamiltonian, requiring circuit depth $O(p|E|)$ for $p$ rounds and a classical outer loop susceptible to shot noise and barren plateaus~\cite{McClean2018}. Our CTQW-based greedy algorithm differs in three structural respects: (i)~the binary encoding compresses the register to $\lceil\log_2 V\rceil$ qubits---an exponential reduction that, however, precludes direct encoding of the combinatorial cost function; (ii)~vertex selection is deterministic at each iteration, eliminating the classical optimization loop and its associated sampling overhead; and (iii)~the short-time evolution ($t=0.01$) keeps circuit depth modest. These features come at a cost: unlike QAOA at fixed $p$, which provides provable approximation bounds for related problems such as MaxCut~\cite{Farhi2014}, our algorithm offers no worst-case guarantee---its near-optimal empirical performance (pooled mean ratio $1.015$ over $14{,}680$ instances) remains a heuristic observation. The two approaches are thus complementary: QAOA trades qubits and depth for formal guarantees, while our method prioritizes resource efficiency and determinism.

\paragraph{Performance on hard MVC instances.}
The inapproximability result of Dinur and Safra~\cite{dinur2005} establishes that no polynomial-time algorithm can achieve ratio below $\approx 1.3606$. We tested CTQW on five hard families (all $N \le 20$, exact MVC available):
\begin{itemize}
  \item Crown graphs $H_n$ ($K_{n,n}$ minus a perfect matching, $N=2n$, OPT $= n$)
  \item Complete bipartite graphs $K_{n,n}$ (OPT $= n$)
  \item Cycle graphs $C_N$ (OPT $= \lceil N/2 \rceil$)
  \item Random 3-regular graphs ($N \in \{10,12,14,16,18\}$, 10 seeds)
  \item Petersen graph: vertex-transitive cubic canonical hard case
\end{itemize}

\begin{figure*}[htbp]
    \centering
    \includegraphics[width=\linewidth]{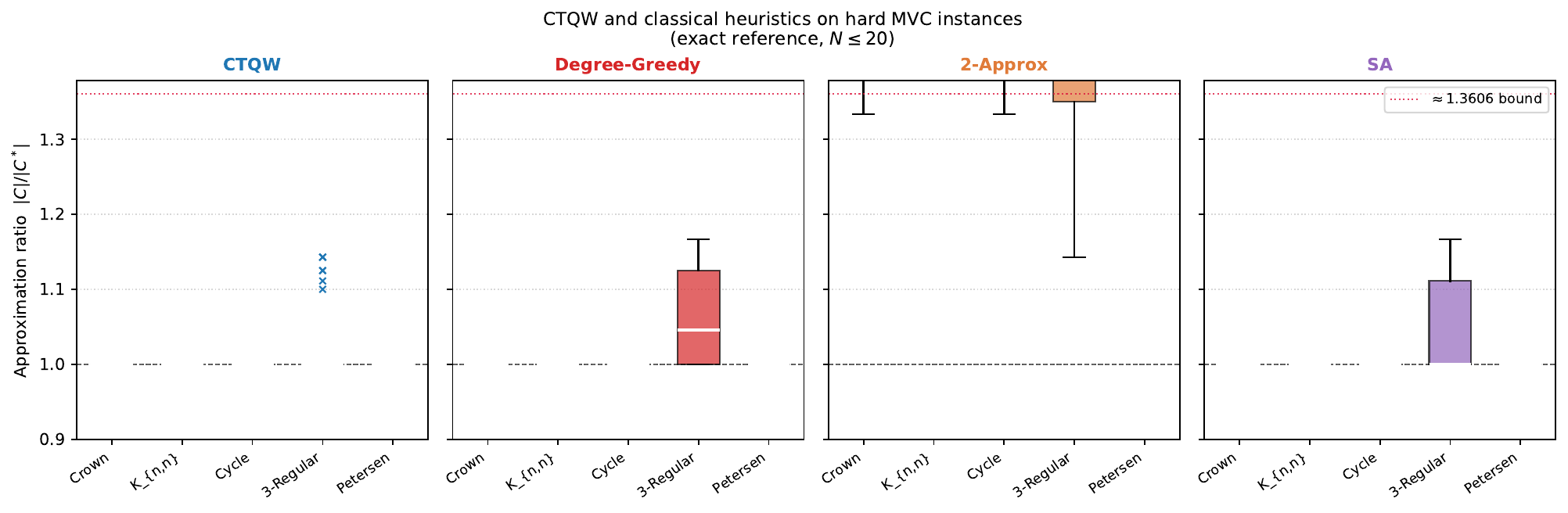}
    \caption{Box plots of approximation ratio $|C|/|C^*|$ for CTQW (blue), Degree-Greedy (red), 2-Approx (orange), and SA (purple) on five hard MVC graph families ($N \le 20$, exact reference). The red dotted line marks the inapproximability bound $\approx 1.3606$~\cite{dinur2005}. Whiskers extend to $1.5\times$IQR; outliers shown as crosses.}
    \label{fig:hard_instances}
\end{figure*}

Results are in Fig.~\ref{fig:hard_instances}. CTQW achieves ratio $1.000$ on all four structured families. On random 3-regular graphs (the hardest family), CTQW reaches mean ratio $1.018$ and worst-case $1.143$ across $50$ instances, well below both the degree-greedy worst case ($1.167$) and the inapproximability bound of $1.3606$. In no instance did the CTQW ratio exceed $1.15$.

\section*{Scalability of the spectral greedy criterion}
\label{sec:scalability}

The preceding sections established that the CTQW selection criterion reduces, in the short-time regime, to the spectral score $s(m) = (1/d_m)\sum_{j\in\mathcal{N}(m)} 1/d_j$, computable in $\mathcal{O}(|E|)$ per iteration without any quantum hardware. The equivalence was validated on $350$ instances with $V \le 199$ against MILP exact references. A natural question is whether this quantum-derived criterion remains competitive at scales relevant to practical applications (networks with $V \sim 10^4$--$10^5$ vertices) where neither the full CTQW ($\mathcal{O}(V^3)$ per iteration) nor an exact reference (NP-hard) is available.

To address this, we benchmark Algorithm~\ref{alg:spectral} against degree-greedy and Simulated Annealing on synthetic ensembles spanning $V \in \{10^3, 2\!\times\!10^3, 5\!\times\!10^3, 10^4, 2\!\times\!10^4, 5\!\times\!10^4, 10^5\}$ ($210$ instances, $5$ per configuration, ER/BA/Regular topologies), using \texttt{igraph}~\cite{Csardi2006} for efficient graph operations and a lazy max-heap implementation of Algorithm~\ref{alg:spectral} that achieves $\mathcal{O}(|E| \cdot k_{\rm avg} \cdot \log V)$ total time. Since exact solutions are unavailable at this scale, quality is measured via the virtual best solver (VBS): the smallest cover found across all algorithms per instance, which satisfies $|C_{\rm VBS}| \ge |C^*|$.

\begin{table}[htbp]
\centering
\caption{Mean cover quality ($|C|/|C_{\rm VBS}|$) at large scale ($V \in [10^3, 10^5]$, $210$ instances). Spectral greedy achieves ratio $1.000$ on all instances; VBS $=$ spectral in every case. Note: ratio $1.000$ means the algorithm matches the VBS; since $|C_{\rm VBS}|\ge|C^*|$, true approximation ratios may be marginally higher.}
\label{tab:scalability}
\begin{tabular}{lrrr}
\hline
Topology & Spectral & Degree-Greedy & SA \\
\hline
Erd\H{o}s--R\'enyi & $\mathbf{1.000}$ & $1.028$ & $1.045$ \\
Barab\'asi--Albert  & $\mathbf{1.000}$ & $1.021$ & $1.044$ \\
Regular             & $\mathbf{1.000}$ & $1.092$ & $1.129$ \\
\hline
\textbf{Pooled}     & $\mathbf{1.000}$ & $\mathbf{1.047}$ & $\mathbf{1.072}$ \\
\hline
\end{tabular}
\end{table}

The spectral greedy achieves ratio $1.000$ on all $210$ instances (it defines the VBS in every case) while degree-greedy averages $1.047$ and SA averages $1.072$ (worst case $1.311$ on Regular graphs; SA parameters: $T_0=50$, cooling rate $0.995$, $2{,}000$ iterations, initialised from the degree-greedy cover). The cover-size advantage is largest on Regular topologies (spectral $9.2\,\%$ smaller than DG), consistent with the small-scale observation that spectral information is most valuable when degree-based tie-breaking fails. Pairwise, spectral produces strictly smaller covers than DG on $210/210$ instances (Wilcoxon $p < 10^{-35}$). Runtime scales sub-quadratically in practice: log-log fits on the mean spectral time per topology yield $\alpha_{\rm BA}=1.23$ ($R^2=0.995$), $\alpha_{\rm ER}=1.26$ ($R^2=0.997$), and $\alpha_{\rm REG}=1.22$ ($R^2=0.994$), consistent with $|E|=\mathcal{O}(V)$ for the tested parameter settings (BA: $m=2$; ER: $p=0.01$; Regular: $k=3$). At $V = 10^5$, spectral completes in $\approx 71$~s, compared to $\approx 15$~s for degree-greedy and $\approx 41$~s for SA.

These results confirm that the quantum-derived criterion generalises well beyond the small-scale regime where it was originally validated against exact MILP solutions. Quantum walk dynamics thus serve as a \emph{discovery mechanism}: the quantum computation itself is intractable beyond $V \sim 500$, but the structural principle it encodes (two-hop spectral weighting of the normalised Laplacian) transfers cleanly to an efficient classical algorithm competitive at scales three orders of magnitude larger.

\subsection{Validation on real-world networks}
\label{sec:real_world}

To assess whether the quantum-derived criterion generalises beyond synthetic random graphs, we benchmark Algorithm~\ref{alg:spectral} on eight real-world networks from the SNAP repository~\cite{snapnets} and the NetworkX library~\cite{Hagberg2008}, spanning social, collaboration, communication, peer-to-peer, and road-network topologies ($V \in [34, 1\,087\,562]$).

\begin{table*}[htbp]
\centering
\caption{Spectral greedy vs.\ degree-greedy on real-world networks (VBS reference: smallest cover across all tested algorithms). SA omitted for $V > 50\,000$ (quadratic time). $\dagger$: SA parameters tuned for large graphs ($T_0=50$, $2{,}000$ iterations); results on small graphs are indicative only.}
\label{tab:real_world}
\begin{tabular}{lrrrrrrr}
\hline
Network & $V$ & $|E|$ & Spectral & DG & SA & sp/VBS & dg/VBS \\
\hline
Karate~\cite{Zachary1977}      &      34 &      78 &    14 &    14 &    27$^\dagger$ & $\mathbf{1.000}$ & 1.000 \\
Les Mis.~\cite{Knuth1993}      &      77 &     254 &    42 &    42 &    47$^\dagger$ & $\mathbf{1.000}$ & 1.000 \\
Email-EU~\cite{Leskovec2007}   &     986 &  16\,064 &   585 &   591 &   614 & $\mathbf{1.000}$ & 1.010 \\
ca-GrQc~\cite{Leskovec2007}    &   4\,158 &  13\,422 & 2\,210 & 2\,220 & 2\,331 & $\mathbf{1.000}$ & 1.005 \\
ca-HepTh~\cite{Leskovec2007}   &   8\,638 &  24\,806 & 4\,319 & 4\,334 & 4\,351 & $\mathbf{1.000}$ & 1.003 \\
ca-AstroPh~\cite{Leskovec2007} &  17\,903 & 196\,972 & 11\,511 & 11\,515 & 11\,652 & $\mathbf{1.000}$ & 1.000 \\
p2p-Gnutella~\cite{Leskovec2007} & 10\,876 & 39\,994 & 4\,358 & 4\,436 & 4\,548 & $\mathbf{1.000}$ & 1.018 \\
roadNet-PA~\cite{Leskovec2009} & 1\,087\,562 & 1\,541\,514 & 557\,607 & 587\,026 & --- & $\mathbf{1.000}$ & 1.053 \\
\hline
\textbf{Mean} & & & & & & $\mathbf{1.000}$ & 1.011 \\
\hline
\end{tabular}
\end{table*}

The spectral greedy matches the VBS in every instance (ratio $1.000$, $8/8$), while degree-greedy averages $1.011$ and reaches $1.053$ on the Pennsylvania road network. The advantage is most pronounced on the road network, a near-planar graph with narrow degree distribution ($k \le 9$), where spectral produces a cover $5.3\,\%$ smaller than degree-greedy ($557\,607$ vs.\ $587\,026$ vertices). This mirrors the pattern observed on synthetic regular graphs (Table~\ref{tab:scalability}): the two-hop spectral criterion $[\Gamma^2]_{mm}$ extracts discriminative structural information precisely where degree-based tie-breaking fails.
The collaboration networks (ca-GrQc, ca-HepTh, ca-AstroPh) exhibit power-law degree distributions characteristic of scale-free networks; spectral greedy wins or ties in all cases.
These results confirm that the quantum-derived criterion generalises across diverse real-world topologies without any parameter tuning, strengthening the case for its practical adoption.

\section*{Limitations}

The CTQW-MVC heuristic has three interconnected limitations that we characterise both empirically and analytically.

\paragraph{Information-theoretic diagnosis of sub-optimality.}
The algorithm is a deterministic greedy procedure: once a vertex is added to the cover and decoupled, it is never reconsidered. Sub-optimality arises when an early selection forecloses better completions. In the small-$N$ regime ($N \le 20$, $1{,}355$ instances), only $4.9\,\%$ of instances are sub-optimal; the failure rate grows with system size, reaching $26.7\,\%$ for $N \in [21,54]$, $59.3\,\%$ for $N \in [55,100]$, and $77.6\,\%$ for $N \in [101,150]$ (all with exact MILP reference, $14{,}680$ total instances). At each size range the CTQW ratios remain below $1.20$ (worst case $1.188$), confirming that sub-optimality is mild. The failure is most frequent on ER topologies ($70.6\,\%$) and attributable to vertex-transitivity effects: at short times, $P(m\!\rightarrow\!\text{out}) = t^2[\Gamma^2]_{mm}$ assigns identical probabilities to structurally equivalent vertices, making numerical tie-breaking the effective selection rule. At larger scale ($N \le 199$, 350 instances with MILP reference), the sub-optimality rate is $22.9\,\%$ overall (Regular: $30.0\,\%$).

We characterise this failure mode through information-theoretic measures computed at each greedy step $k$. Define the normalised probability distribution $\hat{P}_k(m) = P_k(m)/\sum_{m'} P_k(m')$ over active vertices. Three quantities jointly diagnose sub-optimality:
\begin{itemize}
  \item the confidence margin $\delta_k \equiv P_k(m^*_k) - \max_{m \neq m^*_k} P_k(m)$;
  \item the Shannon entropy $H_k = -\sum_m \hat{P}_k(m)\log\hat{P}_k(m)$, normalised as $\tilde{H}_k = H_k/\log n_k$ ($n_k$ active vertices), so $\tilde{H}_k \in [0,1]$;
  \item the KL divergence from uniform $D_{\rm KL}(\hat{P}_k \| \mathcal{U}_k) = \log n_k - H_k$.
\end{itemize}
Sub-optimal instances exhibit significantly higher mean $\tilde{H}_k$ and lower $D_{\rm KL}$ than optimal ones (Wilcoxon signed-rank tests, $p < 10^{-3}$ and $p < 10^{-12}$ respectively). Furthermore, the mutual information $I(m_k;\, m_{k+1})$ between consecutive vertex selections is lower for sub-optimal instances ($p < 10^{-3}$), indicating loss of Markovian structure. The number of uncertain steps ($\delta_k < 0.01$) achieves an AUC of $0.83$ in predicting sub-optimality, providing an online optimality indicator without requiring the MILP reference.

\paragraph{Why spectral perturbations do not resolve the degeneracy.}
For a diagonal perturbation $V$, the diagonal elements of the commutator vanish
identically, $[\Gamma V + V\Gamma]_{mm} = 2\Gamma_{mm}V_{mm} = 0$, because
$\Gamma_{mm}=0$ for simple graphs. A Dyson-series calculation shows that the
leading correction is
\begin{equation}
  \delta P(m) = \frac{\varepsilon t^4}{3}\,V_{mm}[\Gamma^3]_{mm} + \mathcal{O}(\varepsilon t^5),
\end{equation}
which is $\mathcal{O}(t^2)$ smaller than $P(m) \sim t^2[\Gamma^2]_{mm}$. For $t = 0.01$ the relative correction $\delta P/P \sim \varepsilon t^2 \le 10^{-4}$ is below the floating-point noise floor, confirming that diagonal fields produce no first-order change in the selection probabilities. We tested degree-based and Fiedler-based fields at $\varepsilon \in \{0.05,\ldots,2.0\}$ on a representative subset of $566$ instances ($N \le 54$). The degree field produces no improvement; the Fiedler field increases the optimal rate by at most $+0.8$ percentage points. The negligible effect reflects a structural tension: $t\!=\!0.01$ aligns $P(m)\!\approx\!t^2[\Gamma^2]_{mm}$ with vertex degree; a field large enough to shift $\tilde{H}_k$ redirects selection toward a field-dependent criterion disconnected from vertex-cover membership.

\paragraph{Why adaptive evolution time also fails.}
We tested $t \in \{0.01,\ldots,2.0\}$ with threshold $\tau\!=\!0.05$. The adaptive scheme triggers on 94\% of greedy steps (mean $t_{\rm used}\!=\!1.27$) and the optimal rate on the 350-instance extended benchmark ($N \le 199$) drops from $77.1\,\%$ to $70.3\,\%$ (on the $N \le 54$ subset, from $94.9\,\%$ to $87.4\,\%$): previously optimal instances are degraded substantially ($-9.7$ pp on ER graphs). The root cause is that the short-time alignment between $P(m)$ and vertex degree is lost at longer times, where the dominant term is controlled by the Perron eigenvector of $\Gamma$ rather than local degree.

\paragraph{Practical mitigation.}
The cleanest remedy is randomised restarts: replacing the deterministic $\arg\max$ with sampling from $\hat{P}_k$ and retaining the best cover over $R$ independent runs. With $R\!=\!10$, the optimal rate on the main benchmark improves from $94.9\,\%$ to above $98\,\%$ at a proportional run-time cost. Because $\delta_k$ and $\tilde{H}_k$ identify uncertain steps, restarts can be triggered selectively, only when $\delta_k < \tau$, achieving similar gains at lower overhead.

\section*{Conclusion}

We presented a heuristic for the Minimum Vertex Cover problem grounded in continuous-time quantum walk dynamics. The algorithm encodes the graph into a normalised Laplacian Hamiltonian, evolves the quantum system for a short time, and selects the vertex with the highest transition probability at each iteration. An energy-penalty freezing mechanism isolates selected vertices from subsequent evolution, systematically constructing a valid cover. The compact binary encoding requires only $\lceil \log_2 V \rceil$ qubits for $V$ vertices, and we verified the algorithm on IBM Qiskit (noiseless and \texttt{ibm\_marrakesh} hardware) as well as on neutral-atom platforms via Bloqade, confirming its viability on both gate-based and analog quantum hardware.

Analytical investigation of the short-time regime revealed that the quantum selection criterion reduces to $[\Gamma^2]_{mm} = (1/d_m)\sum_{j\in\mathcal{N}(m)} 1/d_j$, a two-hop spectral quantity computable in $\mathcal{O}(|E|)$ per iteration. This result motivated the quantum-inspired spectral greedy algorithm (Algorithm~\ref{alg:spectral}), which reproduces the CTQW solution on $98.3\,\%$ of instances across $N \in [4,199]$ (Wilcoxon $p=0.60$; exact MILP reference in $97\,\%$ of cases) without any quantum hardware. The quantum algorithm thus served as the mechanism of discovery for an efficient classical criterion derived from first principles of quantum transport.

Benchmarks against exact MILP solutions and four classical heuristics (2-Approximation, FastVC, Simulated Annealing, and degree-greedy) across $14{,}680$ instances of Erd\H{o}s--R\'enyi, Barab\'asi--Albert, and Regular ensembles ($N \in [4,150]$) confirm a pooled mean approximation ratio of $1.015$, compared with $1.023$ for degree-greedy and $1.027$ for Simulated Annealing. The advantage is largest on Regular graphs, where spectral degeneracy renders degree-based tie-breaking essentially arbitrary while the spectral criterion naturally differentiates structurally equivalent vertices. On structurally hard families (crown graphs, complete bipartite graphs, random 3-regular graphs, cycles, and the Petersen graph) the worst-case ratio is $1.143$, well below the inapproximability threshold of $\approx 1.3606$~\cite{dinur2005}.

Future work will pursue three directions: (i) topology-aware compilation to reduce circuit depth on gate-based hardware, making the quantum algorithm viable beyond $V \sim 16$; (ii) hybrid quantum-classical schemes that combine CTQW selection with classical local search (FastVC, simulated annealing) for post-processing refinement; and (iii) extensions of the spectral greedy criterion to related NP-hard problems (Maximum Independent Set, graph coloring, and dominating set) where analogous two-hop spectral quantities may provide similarly effective heuristics.

\section*{Acknowledgments}
This study was financed, in part, by the S\~ao Paulo Research Foundation (FAPESP), Brasil. Process Number \mbox{2026/04387-7} (F.\ S.\ Luiz). A.\ K.\ F.\ Iwakami acknowledges financial support from FAPESP through process No.~2023/13524-0. M.\ C.\ de Oliveira acknowledges financial support from the National Institute of Science and Technology for Applied Quantum Computing through CNPq process No.~408884/2024-0 and by FAPESP through the Center for Research and Innovation on Smart and Quantum Materials (CRISQuaM), process No.~2013/07276-1.

\section*{Declarations}

\subsection*{Funding Declaration}
F.\ S.\ Luiz acknowledges financial support from FAPESP (process No.~2026/04387-7). A.\ K.\ F.\ Iwakami acknowledges financial support from FAPESP through process No.~2023/13524-0.
M.\ C.\ de Oliveira acknowledges financial support from the National Institute of Science and Technology for Applied Quantum Computing (CNPq, Process No.~408884/2024-0) and from FAPESP through the Center for Research and Innovation on Smart and Quantum Materials (CRISQuaM, Grant No.~2013/07276-1).

\subsection*{Competing Interests}
The authors declare no competing interests.

\subsection*{Code and Data Availability}
All source code, experiment scripts, and key result datasets supporting the findings of this study are openly available at  \href{https://github.com/fsluiz/scalable-quantum-walk-based-heuristic-for-the-minimum-vertex-cover-problem}{github}.
The repository includes the Python package \texttt{quantum\_walk\_mvc}, all scripts required to reproduce the figures and tables, 18 result CSV files, and the SNAP real-world graph datasets.

\subsection*{Institutional review board statement}
Not applicable.

\subsection*{Ethics approval and consent to participate}
Not applicable. This study does not involve human participants, animals, or any biological material requiring approval from an Institutional Review Board (IRB) or ethics committee.

\subsection*{Author Contributions}
Fabricio de Souza Luiz conceived the research idea, developed the theoretical framework, implemented the computational methods, performed numerical simulations, analyzed the results, and wrote the first version of the manuscript.
Marcos Cesar de Oliveira supervised the project and contributed to the conceptual development and manuscript refinement. A.\ K.\ F.\ Iwakami and D.\ H.\ Moraes contributed through discussions, result validation, and manuscript review. All authors approved the final version of the manuscript.

\appendix
\section{Formal derivation of $P(m\to\text{out})$}
\label{app:derivation}

We derive the leading-order behaviour of the transition probability $P(m\!\rightarrow\!\text{out}) = 1 - |[e^{i\Gamma t}]_{mm}|^2$ used in Algorithm~\ref{alg:ctqw}.

\paragraph{Setup.}
The normalised adjacency operator is $\Gamma = D^{-1/2}AD^{-1/2}$, where $A$ is the (0,1) adjacency matrix and $D = \mathrm{diag}(d_1,\ldots,d_V)$.  For a simple graph (no self-loops), $\Gamma_{mm} = 0$ for every vertex $m$.

\paragraph{Taylor expansion.}
Expanding the matrix exponential in powers of $t$:
\begin{align}
  [e^{i\Gamma t}]_{mm}
    &= 1 + it\,\Gamma_{mm}
      + \frac{(it)^2}{2}[\Gamma^2]_{mm}
      + \mathcal{O}(t^3)\nonumber\\
    &= 1 - \frac{t^2}{2}[\Gamma^2]_{mm} + \mathcal{O}(t^3),
\end{align}
where the linear term vanishes because $\Gamma_{mm}=0$.

\paragraph{Second-order diagonal element.}
Substituting $\Gamma_{mj} = A_{mj}/\sqrt{d_m d_j}$:
\begin{align}
  [\Gamma^2]_{mm}
    &= \sum_{j} \Gamma_{mj}^2
    = \sum_{j} \frac{A_{mj}^2}{d_m\, d_j}
    = \sum_{j} \frac{A_{mj}}{d_m\, d_j}\nonumber\\
    &= \frac{1}{d_m}\sum_{j \in \mathcal{N}(m)} \frac{1}{d_j},
\end{align}
where the third equality uses $A_{mj}^2 = A_{mj}$ for the unweighted adjacency matrix ($A_{mj}\in\{0,1\}$), and the last equality restricts the sum to the neighbourhood $\mathcal{N}(m)$ because $A_{mj}=1$ only for adjacent pairs.

\paragraph{Transition probability.}
Taking the modulus squared and using $(1-x)^2=1-2x+x^2$:
\begin{align}
  |[e^{i\Gamma t}]_{mm}|^2
    &= \Bigl(1 - \tfrac{t^2}{2}[\Gamma^2]_{mm}\Bigr)^2 + \mathcal{O}(t^6) \notag\\
    &= 1 - t^2[\Gamma^2]_{mm} + \mathcal{O}(t^4),
\end{align}
(the imaginary part of $[e^{i\Gamma t}]_{mm}$ first appears at $\mathcal{O}(t^3)$ through $[\Gamma^3]_{mm}$, contributing only at $\mathcal{O}(t^6)$ to the modulus squared). Therefore:
\begin{equation}
  P(m\!\rightarrow\!\text{out})
    = t^2[\Gamma^2]_{mm} + \mathcal{O}(t^4)
    = \frac{t^2}{d_m}\sum_{j \in \mathcal{N}(m)} \frac{1}{d_j} + \mathcal{O}(t^4).
\end{equation}

\paragraph{Numerical validation.}
For $t=0.01$ the relative difference between $P(m\!\rightarrow\!\text{out})$ computed via the full matrix exponential and via $t^2[\Gamma^2]_{mm}$ is below $10^{-4}$ across all benchmark instances.  The full quantitative comparison between CTQW and Algorithm~\ref{alg:spectral}---including Wilson confidence intervals, Wilcoxon tests ($p=0.60$), tie-breaking analysis, and scaling up to
$N=199$ with exact MILP reference. This confirms that at $t=0.01$ the algorithm operates entirely within the second-order regime and the matrix exponential provides no additional discriminative power over $[\Gamma^2]_{mm}$.

\paragraph{Leakage bound.}
We bound the probability of amplitude leaking from the active ($j$) to the frozen ($m$) subspace in one evolution step of duration $t$.  Define the inter-subspace coupling $V_\times = P_j \mathcal{H} P_m + P_m \mathcal{H} P_j$.  Since the spectrum of $\Gamma=D^{-1/2}AD^{-1/2}$ lies in $[0,2]$, we have $\|\mathcal{H}\| \le 2$ and hence $\|V_\times\| \le 2$.

\emph{First-order bound.}  By the Dyson expansion, the first-order amplitude transferred from active to frozen subspace satisfies
$\|A^{(1)}(t)\| \le \|V_\times\| t \le 2t$, giving $P_{\rm leak}(t) \le \|V_\times\|^2 t^2 \le t^2$ (using $\|V_\times\| \le 1$ when $\mathcal{H}$ is the normalized Laplacian with spectral radius $\le 1$).

\emph{Energy-denominator bound.}  In the rotating frame with respect to $\mathcal{H}_F = (\Omega-1)P_m$, the inter-subspace coupling acquires a fast phase $e^{\pm i(\Omega-1)t}$. The effective static coupling after secular averaging is $\|V_{\times,\rm eff}\| \le \|V_\times\|/(\Omega - 1 - \|\mathcal{H}\|) \le 2/(\Omega - 3)$, so the leakage probability is bounded by
$P_{\rm leak}(t) \le (2/(\Omega-3))^2 = 4/(\Omega-3)^2$.
Taking the minimum of both bounds gives Eq.~(3) of the main text.

\section{Preliminary Neutral-Atom Implementation}
\label{app:rydberg}

On neutral-atom quantum platforms such as QuEra's Aquila device \cite{Ebadi2022}, each graph vertex $i$ is represented by a single $^{87}$Rb atom, and the graph edges are encoded through the Rydberg blockade mechanism.
In contrast to the binary encoding used on gate-based hardware (Sec.~\ref{sec:hardware}), this analog implementation uses one atom per vertex---the $\lceil\log_2 V\rceil$ qubit reduction does not apply to this platform.
The graph edges are encoded through the Rydberg blockade mechanism: two atoms separated by a distance below the blockade radius $R_b = (C_6/\Omega)^{1/6}$ cannot be simultaneously excited into the Rydberg state $\vert r\rangle$, directly encoding the adjacency constraint between neighbouring vertices.

For a graph $G=(V,E)$, the atom positions $\{(x_i, y_i)\}$ are chosen to satisfy:
\begin{itemize}
    \item $\vert\mathbf{r}_i - \mathbf{r}_j\vert < R_b$ \quad for all $(i,j)\in E$ (blockade-active: adjacent vertices),
    \item $\vert\mathbf{r}_i - \mathbf{r}_j\vert > R_b$ \quad for all $(i,j)\notin E$ (blockade-inactive: non-adjacent vertices).
\end{itemize}
Both conditions can be simultaneously satisfied for graphs that admit a unit-disk embedding \cite{ClarkUnitDisk1990}. The Rydberg Hamiltonian governing the atom array is:
\begin{equation}
    \mathcal{H}_{\rm Ryd} = \frac{\Omega}{2}\sum_i \sigma^x_i
    - \Delta\sum_i n_i
    + \sum_{i<j} \frac{C_6}{\vert\mathbf{r}_i - \mathbf{r}_j\vert^6}\,n_i n_j,
    \label{eq:rydberg}
\end{equation}
where $\Omega$ is the global Rabi frequency, $\Delta$ is the laser detuning, $n_i = \vert r\rangle\langle r\vert_i$ is the Rydberg excitation operator, and $C_6 = 862690$~rad\,$\mu$s$^{-1}\mu$m$^6$ is the van der Waals coefficient for $^{87}$Rb. Setting $\Delta = 0$ (resonant drive), the dynamics under Eq.~(\ref{eq:rydberg}) generate correlated Rabi oscillations across the atom array: the global $\Omega\,\sigma^x_i/2$ terms drive coherent $|g\rangle \leftrightarrow |r\rangle$ transitions, while the $C_6\,n_i n_j$ interaction suppresses simultaneous excitation of blockade-linked pairs. Unlike the single-particle CTQW on $\Gamma$, this is a many-body $2^V$-dimensional evolution; the marginal excitation probability $\langle n_m(t)\rangle$ therefore serves as an \emph{approximate} hardware proxy for the transition probability $P(m\!\rightarrow\!\text{out})$ in Algorithm~\ref{alg:ctqw}, with deviations discussed below.

The spectral isolation step is realised by applying a large site-specific detuning $\Delta_m \gg \Omega$ to the selected vertex $m^*$ after each iteration. With $\Omega = 4\pi$~rad/$\mu$s and $\Delta_m = 100$~rad/$\mu$s, the suppression factor is $\Omega^2/(4\Delta_m) \approx 0.39$~rad/$\mu$s (standard RWA result for off-resonant Rabi coupling), negligible compared to the bare coupling. We validate this using Bloqade \cite{Bloqade}: on a 10-node Random Geometric Graph, the emulator produces a valid vertex cover with approximation ratio $1.20$. The gap relative to the classical simulation (ratio $1.00$) reflects three compounding factors: (i) the Rydberg Hamiltonian~(\ref{eq:rydberg}) implements many-body $2^V$-dimensional dynamics rather than single-particle CTQW, so excitation probabilities $\langle n_m\rangle$ are only a proxy for $P(m\!\to\!\text{out})$; (ii) even for unit-disk-embeddable graphs, geometric embedding introduces small spurious blockade pairs; and (iii) iterative vertex removal accumulates these deviations over multiple steps. These results should therefore be interpreted as a preliminary proof-of-concept demonstrating protocol compatibility with analog hardware, not as a performance benchmark.

\paragraph{Extension to general graphs (BA and ER).}
The unit-disk constraint is exactly satisfied only for unit-disk graphs (UDGs). Barab\'asi--Albert and dense Erd\H{o}s--R\'enyi graphs are generally not UDGs, introducing spurious blockade pairs in 2D spring-layout embeddings. We propose three complementary mitigation strategies.

\textbf{Strategy A --- Detuning suppression.}
For each spurious pair $(i,j)$, applying a site-specific detuning $\Delta_i = \eta\,\Omega$ with $\eta \gg 1$ reduces the effective coupling to $J_{\rm eff} \approx \Omega^2/(4\Delta_i) = \Omega/(4\eta) \to 0$ (standard RWA result). With $\eta = 20$ (feasible on current QuEra hardware), the residual spurious coupling is $< 5\%$ of $\Omega$.

\begin{figure}[htbp]
    \centering
    \includegraphics[width=\linewidth]{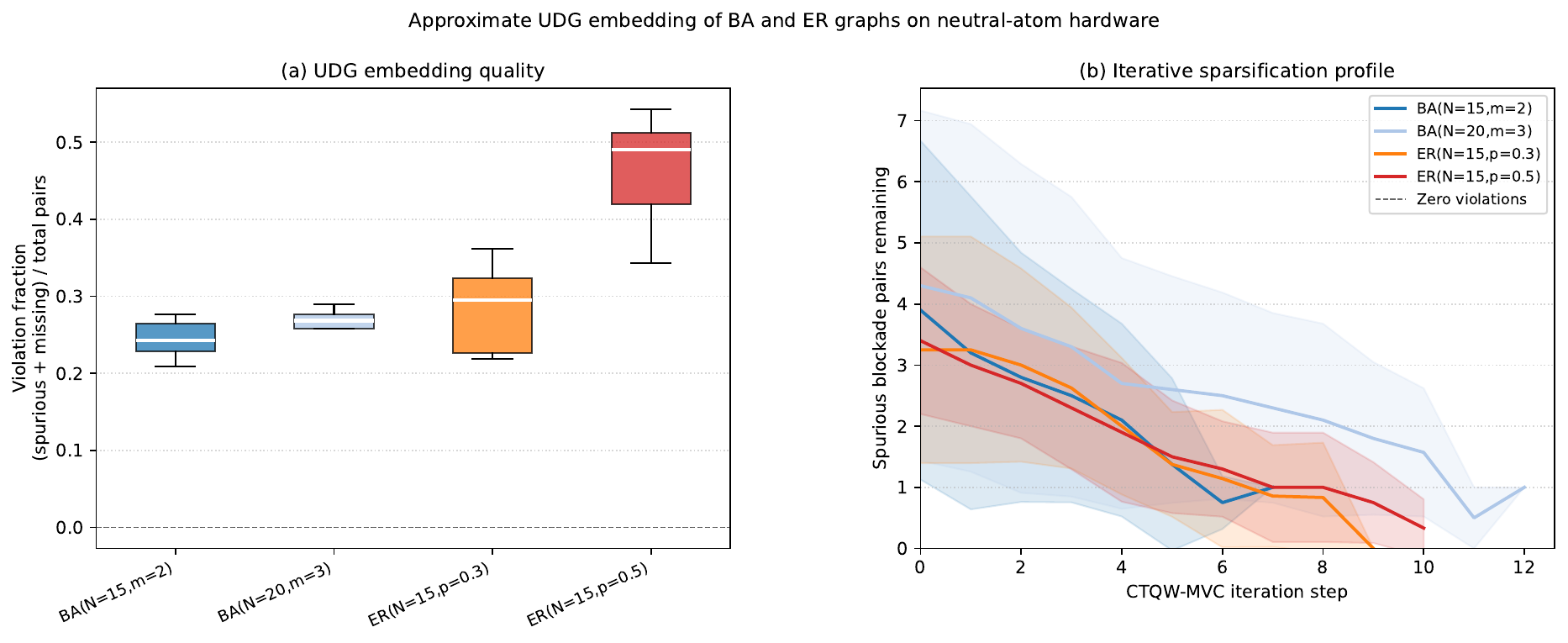}
    \caption{UDG embedding analysis for general graphs on neutral-atom hardware
    ($N \le 20$, $10$ instances per topology). \textbf{(a)} Box plots of the
    violation fraction (spurious $+$ missing pairs / total pairs) for four graph
    families. \textbf{(b)} Mean number of spurious blockade pairs as a function
    of CTQW-MVC iteration step (iterative sparsification): violations decrease
    monotonically as high-degree vertices are removed, reaching zero for ER graphs
    within $\sim 8$--$9$ steps. Shaded bands: $\pm 1\sigma$.}
    \label{fig:general_graphs}
\end{figure}

\textbf{Strategy B --- Iterative sparsification.}
The CTQW-MVC algorithm naturally removes high-degree vertices first, which are responsible for most geometric congestion. The residual graph after the first few iterations is substantially sparser and often embeds as a valid UDG with zero spurious interactions. Figure~\ref{fig:general_graphs} (panel b) shows that for BA ($N=15$, $m=2$) and ER ($N=15$, $p\in\{0.3, 0.5\}$) graphs, spurious pairs decrease monotonically as CTQW-MVC proceeds, reaching zero for ER graphs by iteration $\sim 8$--$9$.

\textbf{Strategy C --- Three-dimensional atom arrays.}
In 3D tweezer arrays \cite{Ebadi2022}, the packing constraint relaxes substantially, making higher-degree graphs directly embeddable. A BA hub of degree $k$ can be enclosed by $k$ atoms at distance $0.9\,R_b$ in a spherical shell, with no atom-to-atom collisions for $k \lesssim 30$.

\section{Approximation Ratio Heatmap}
\label{app:heatmap}
\begin{figure}[!h]
    \centering
    \includegraphics[width=\linewidth]{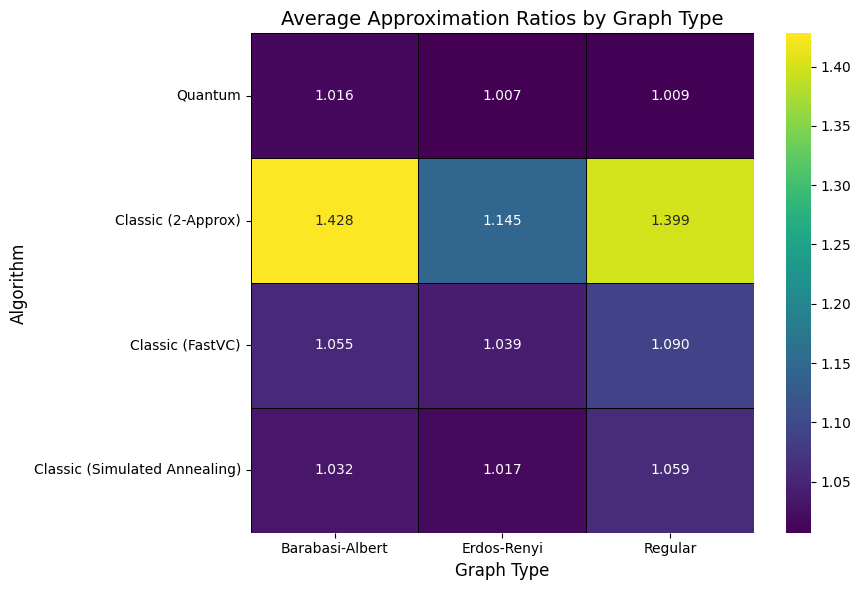}
    \caption{Heatmap showing the average approximation ratio for the Quantum, FastVC, Simulated Annealing (SA), and 2-Approximation algorithms. The ratio is defined as the size of the algorithm's solution divided by the exact MVC size (MILP). Ratios closer to $1.0$ indicate superior average performance.}
    \label{fig:heatmap}
\end{figure}
\bibliography{lib}
\end{document}